\documentclass[%
 reprint,
 amsmath,amssymb,
 aps,
]{revtex4-1}

\usepackage{graphicx}
\usepackage{bm}
\usepackage{braket}

\newcommand{\average}[1]{\ensuremath{\langle#1\rangle} }
\def\vector#1{\mbox{\boldmath $#1$}}

\begin{document}

\preprint{APS/123-QED}

\title{Multiorbital Ferroelectric Superconductivity in doped SrTiO$_3$}

\author{Shota Kanasugi}
\thanks{kanasugi.shouta.62w@st.kyoto-u.ac.jp}
\author{Youichi Yanase}%
\affiliation{%
 Depertment of Physics, Graduate School of Science, Kyoto University, Kyoto 606-8502, Japan
}%
\date{\today}

\begin{abstract}
SrTiO$_3$ is a unique example of a system which exhibits both quantum paraelectricity and superconductivity. 
Thus, it is expected that the superconducting state is closely related to the intrinsic ferroelectric instability. 
Indeed, recent experiments suggest existence of a coexistent phase of superconductivity and ferroelectricity in Ca-substituted SrTiO$_3$. 
In this paper, we propose that SrTiO$_3$ can be a platform of the ferroelectric superconductivity, which is characterized by a ferroelectric transition in the superconducting state. 
By analyzing a multiorbital model for $t_{2g}$ electrons, we show that the ferroelectric superconductivity is stabilized through two different mechanisms which rely on the presence of the spin-orbit coupling. 
First, the ferroelectric superconducting state is stabilized in the dilute carrier density regime due to a ferroelectricity-induced Lifshitz transition. 
Second, it is stabilized under a magnetic field independent of the carrier density. 
The importance of the multiorbital or multiband nature for the ferroelectric superconductivity is clarified. 
Then, we predict a topological Weyl superconducting state in the ferroelectric superconducting phase of SrTiO$_3$.
\end{abstract}

\pacs{Valid PACS appear here}
\maketitle


\section{\label{sec:Intro}Introduction}
The origin of the superconductivity in SrTiO$_3$ (STO) has remained to be a long standing problem of the condensed matter physics for more than half a century. 
The superconductivity in doped STO begins to emerge at an extraordinary low carrier density on the order of 10$^{17}$ cm$^{-3}$ \cite{PhysRevX.3.021002,PhysRevLett.112.207002}, where the Fermi energy $\epsilon_{\rm F}$ is smaller than the characteristic phonon energy $\omega_{\rm D}$. 
Therefore, conventional BCS or Migdal-Eliashberg theories are invalid for this superconducting state since the retardation condition ($\epsilon_{\rm F}\gg\omega_{\rm D}$) is not satisfied. 
To reveal the origin of the dilute superconducting state, various pairing glues have been proposed theoretically, e.g., plasmons \cite{takada1980theory,PhysRevB.94.224515}, localized longitudinal optical phonons \cite{Gorkov4646}, and soft transverse optical phonons \cite{PhysRevLett.115.247002,PhysRevB.97.144506,PhysRevB.98.104505,PhysRevMaterials.2.104804,PhysRevB.98.220505}. 
However, there is still no consensus about the pairing mechanism in STO. 

Recently, the superconductivity in STO receives extensive attention also in terms of the ferroelectric (FE) quantum criticality. 
STO is a quantum paraelectric (PE) \cite{PhysRevB.19.3593} which exists in the vicinity of the FE quantum critical point \cite{rowley2014ferroelectric}. 
The avoided FE ordering can be activated by some chemical or physical operations, e.g., isovalent substitution of Sr with Ca \cite{PhysRevLett.52.2289}, isotopic substitution of $^{16}$O with $^{18}$O \cite{PhysRevLett.82.3540}, and application of tensile strain \cite{PhysRevB.13.271} or electric field \cite{PhysRevB.52.13159}. 
Incidentally, doped STO exhibits metallic behavior at very low carrier densities on the order of 10$^{16}$ cm$^{-3}$\cite{uwe1985conduction,PhysRevX.3.021002,bhattacharya2016spatially} thanks to the quantum paraelectricity and resulting long effective Bohr radius \cite{PhysRevB.17.2575}. 
Thus, it is naturally considered that the dilute superconductivity and the FE quantum criticality are closely related. 
Indeed, enhancement of the superconducting transition temperature $T_{\rm c}$ by a FE quantum fluctuation was proposed theoretically \cite{PhysRevLett.115.247002}, and later confirmed experimentally \cite{stucky2016isotope,NatPhys.13.643-648,tomioka2019enhanced,herrera2018strain,ahadi2019enhancing}. 
Furthermore, a phase transition structurally indistinguishable from the FE phase transition was observed in metallic Sr$_{1-x}$Ca$_{x}$TiO$_{3-\delta}$ \cite{NatPhys.13.643-648} similarly to a FE metal LiOsO$_3$ \cite{NatMater.12.1024-1027}. 
This experimental observation suggests existence of the superconducting phase which coexists with the ferroelectricity. 
Therefore, STO is a candidate of a FE superconductor in which FE-like phase transition occurs in the superconducting state. 

Another extensively debated issue about the superconducting STO is its multiband nature. 
Early tunneling measurements on doped STO observed two peaks in the local density of states (DOS) \cite{PhysRevLett.45.1352} which implies the multiple superconducting gaps. 
This result is supported by recent quantum oscillation measurements \cite{PhysRevLett.112.207002} and thermal conductivity data \cite{PhysRevB.90.140508}. 
Thus, it has been suggested that STO is a multiband superconductor with multiple nodeless gaps, and the multiband effect has been theoretically discussed \cite{PhysRevB.87.014510,edge2015upper,PhysRevLett.121.127002}. 
In contrast, recent tunneling experiments \cite{swartz2018polaronic} and optical conductivity data \cite{PhysRevLett.120.237002} indicate only single superconducting gap. 

\begin{figure}[tbp]
 \centering
   \includegraphics[width=85mm]{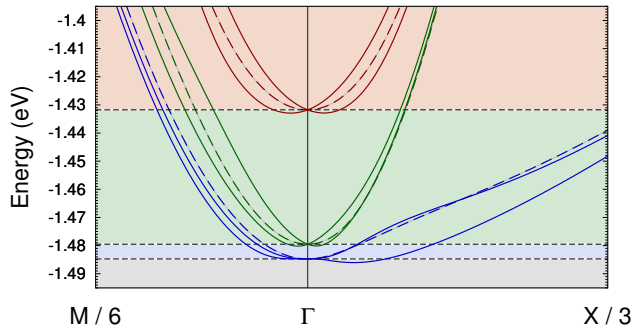}
   \caption{Band structure of bulk STO around the $\Gamma$-point in the PE phase (dashed line; $\gamma=0$) and the FE phase (solid line; $\gamma=27.7$ meV). The Lifshitz transitions occur when the Fermi level crosses black dashed lines. 
Different colored areas show different carrier density regimes which are distinguished by the topology of Fermi surfaces. \label{fig:Band}}
\end{figure}

Although multiband nature in the superconducting STO is still under debate, it is certainly true that the superconducting state has multiorbital features. 
The conduction bands in STO originate from three Ti $t_{2g}$ orbitals. 
Three-fold degeneracy of the $t_{2g}$ orbitals is lifted by the spin-orbit coupling and the tetragonal crystal field due to antiferrodistortive (AFD) rotation of TiO$_6$ octahedra below 105 K \cite{hayward1999cubic}. 
Thus, STO has three distinct bands all centered at the $\Gamma$-point and constructed from multiple orbitals [Fig. \ref{fig:Band}]. 
Therefore, the multiorbital features may affect superconductivity even in the dilute carrier density regime with single Fermi surface. 
Consequently, the superconductivity in doped STO has multiorbital character regardless of the carrier density. 

Considering all the unique aspects of the superconducting STO, in this paper, we show that STO can be a platform of the FE superconductivity through two different mechanisms that rely on the antisymmetric spin-orbit coupling (ASOC). 
First mechanism originates from the ferroelectricity-induced Lifshitz transition in dilute carrier density regimes. 
Another one is the magnetic-field-induced FE superconductivity caused by spin-momentum locking in the FE phase. 
It is shown that, in both mechanisms, the FE superconductivity is strongly influenced by the multiorbital or multiband nature of STO. 
In particular, we predict a Weyl FE superconducting state arising from the multiorbital effect. 

This paper is constructed as follows. 
In Sec. \ref{sec:Model}, a three-orbital model describing the electronic structure and superconductivity of bulk STO is introduced. 
We analyze the model within the mean-field BCS theory. 
The energy of polar lattice distortion is also included phenomenologically to discuss the FE-like structural phase transition. 
In Sec. \ref{sec:Multiorbital}, we discuss the multiorbital effect in STO. 
It is shown that the electronic structure and the Pauli depairing effect of the FE STO are strongly affected by the unconventional Rashba spin-orbit coupling due to the multiorbital effect. 
In addition, a cooperative relationship between multiple Lifshitz transitions in multiorbital electronic structure and FE superconductivity is clarified. 
In Sec. \ref{sec:FESC}, we show the phase diagrams for the FE superconductivity in three different carrier density regimes in STO. 
It is demonstrated that the FE superconducting state is stabilized at zero magnetic field only in a dilute carrier density regime. 
On the other hand, the FE superconducting state is stabilized under the magnetic field irrespective of the carrier density. 
Furthermore, we show that a topological Weyl FE superconducting state is stabilized in a dilute density regime, thanks to  the multiorbital effect. 
Finally, a brief summary and conclusion are given in Sec. \ref{sec:Summary}. 

\section{\label{sec:Model}Model and formulation}
\subsection{\label{sec:t2g_model}Three-orbital model for bulk SrTiO$_3$}
In order to describe the three distinct band structure in tetragonal STO, we introduce a three-orbital tight-binding model for $t_{2g}$ electrons as follows: 
\begin{align}
\mathcal{H}_0 &= 
\sum_{{\bm k},l,\sigma} \left( \varepsilon_l({\bm k}) - \mu \right)  c_{{\bm k}, l \sigma}^{\dag}c_{{\bm k}, l \sigma} +
\lambda  \sum_{{\bm i}} {\bm L}_{{\bm i}} \cdot {\bm S}_{{\bm i}}, 
\label{eq:H_t2g_STO}
\end{align}
where $c_{\bm{k},l\sigma}$ is the annihilation operator for an electron with momentum $\bm{k}$, orbital $l=yz,xz,xy$, and spin $\sigma=\uparrow, \downarrow$. 
The first term is the kinetic-energy term of $t_{2g}$ orbitals including the chemical potential $\mu$. 
The single electron kinetic energy $\varepsilon_{l}({\bm k})$ are described as 
\begin{eqnarray}
\varepsilon_{yz}({\bm k})=&&-2 t_1 \left(\cos k_y + \cos k_z \right)\nonumber\\
&&-2 t_2 \cos k_x-4 t_3 \cos k_y \cos k_z, \\
\varepsilon_{xz}({\bm k})=&& -2 t_1 \left(\cos k_x + \cos k_z\right)\nonumber\\
&&-2 t_2 \cos k_y -4 t_3 \cos k_x \cos k_z, \\
\varepsilon_{xy}({\bm k})=&&-2 t_1 \left(\cos k_x + \cos k_y\right)\nonumber\\
&&-2 t_2 \cos k_z -4 t_3 \cos k_x \cos k_y + \Delta_{\rm T}.
\end{eqnarray}
Here, $\Delta_{\rm T}$ expresses the tetragonal crystal field for the AFD transition, which lifts the energy of the $d_{xy}$ orbital. 
Although the intersite hybridization term has been generally considered for perovskite oxides \cite{yanase2013electronic} in addition to the above three terms, first-principles band calculations have shown that it is negligible in the bulk STO \cite{hirayama2012ab,PhysRevB.86.125121,PhysRevB.87.161102}. 
The second term of Eq. (\ref{eq:H_t2g_STO}) represents the LS coupling of Ti ions. 
The band structure of tetragonal STO is reproduced by $\mathcal{H}_{0}$ with the parameter set listed in Table \ref{tab:STO_parameters}, which is determined based on the first principles calculations \cite{hirayama2012ab,PhysRevB.86.125121,PhysRevB.87.161102}. 

\begin{table}[tbp]
\centering
\caption{\label{tab:STO_parameters}
Model parameters for bulk STO. We choose the unit of energy as $t_{1}=1$. The values of $\Delta_{\rm T}$ and $\lambda$ are set to be larger than the literature values \cite{hirayama2012ab,PhysRevB.86.125121,PhysRevB.87.161102} for simplicity of the numerical calculations. The value of $\gamma$ at the SrTiO$_3$/LaAlO$_3$ interface \cite{PhysRevB.87.161102} is also shown for reference. }
\begin{ruledtabular}
\begin{tabular}{ccc}
  & Literature values \cite{hirayama2012ab,PhysRevB.86.125121,PhysRevB.87.161102}  & This paper \\ \hline
 $t_1$ & 277 meV & 1 \\
 $t_2$ & 31 meV & 0.11 \\
 $t_3$ & 76 meV & 0.27 \\
 $\Delta_{\rm T}$ & 3.2 meV & 0.03 \\
 $\lambda$ & 12 meV & 0.12 \\
 $\gamma$ & 20 meV (SrTiO$_3$/LaAlO$_3$) & $\lesssim$ 0.20 
\end{tabular}
\end{ruledtabular}
\end{table}

Then, we discuss effects of the ferroelectricity on the electronic structure. 
Since the electric polarization is not well defined in metallic or superconducting states, we define the ferroelectricity in conducting systems as a spontaneous nonpolar-to-polar inversion symmetry breaking. 
This FE transition is realized by opposite displacement of Sr/Ti cation and O anion, and thus the crystal symmetry descends to one of polar space groups. 
Although STO has two FE modes parallel and perpendicular to the AFD rotation axis \cite{aschauer2014competition}, we only consider the former for simplicity. 
Thus, the crystallographic space group of tetragonal STO descends to $I4cm$ ($C_{4v}^{10}$) from $I4/mcm$ ($D_{4h}^{18}$) as a consequence of the FE ordering along the [001] axis \cite{watanabe2019accurate}. 
In this mirror symmetry broken FE phase, an orbital hybridization term 
\begin{equation}
\mathcal{H}_{\rm pol} = \sum_{{\bm k},\sigma} \sum_{l=yz,xz} \left[\zeta_{l}(\bm{k}) c_{\bm{k},l\sigma}^{\dag} c_{\bm{k},xy\sigma} + \rm{H.c.} \right] , \label{eq:H_pol} 
\end{equation}
is induced in addition to $\mathcal{H}_0$ \cite{PhysRevB.86.125121}. 
Here, $\zeta_{yz,xz}(\bm{k})=2i\gamma\sin k_{x,y}$. 
Equation (\ref{eq:H_pol}) describes the intersite hybridization between $d_{xy}$ and $d_{yz,xz}$ orbitals, which have different mirror parity along the [001] axis. 
Combination of $\mathcal{H}_{\rm pol}$ and the LS coupling leads to the Rashba ASOC \cite{yanase2007magnetic,PhysRevB.87.161102}, and thus spin-orbit splitting in the band structure is induced in the FE phase [Fig. \ref{fig:Band}]. 
Since this orbital parity mixing is considered to be enhanced under a large polar crystal field \cite{PhysRevB.86.125121}, we assume that the odd-parity hopping integral $\gamma$ is proportional to the polar lattice displacement $P$, i.e., $\gamma\propto P$. 
Hence, we treat $\gamma$ as an order parameter which characterizes the ferroelectricity in STO.

Although the origin of superconductivity in STO is unclear, thermodynamic properties such as the specific heat jump \cite{PhysRevB.90.140508} are in good agreement with the BCS theory. 
Therefore, we investigate an interplay of superconductivity and ferroelectricity by adopting a simple BCS-type model, and focus on the multiorbital effect on the FE superconductivity in STO. 
More precise studies including a realistic dynamical electron-phonon coupling and Coulomb interactions are left for a future work. 
The BCS-type static attractive interaction is introduced as follows: 
\begin{align}
\mathcal{H}_{\rm pair} &= -\frac{V_{s}}{N} 
\sum_{{\bm k},{\bm k}',{\bm q},l}
c_{{\bm k}, l \uparrow}^{\dag} c_{-{\bm k}+{\bm q}, l \downarrow}^{\dag}
c_{-{\bm k}'+{\bm q}, l \downarrow} c_{{\bm k}', l \uparrow}, 
\label{eq:H_pair_STO}
\end{align}
where $N$ is the number of Ti sites, and $\bm{q}$ is the center-of-mass momentum of Cooper pairs. 
Since the $s$-wave superconductivity in STO has been confirmed \cite{PhysRevB.90.140508}, we assume momentum-independent intraorbital pairing interaction. 
The pairing interaction strength $V_s$ is determined to satisfy $T_{\rm c} \ll E_{\rm SO}$, where $T_{\rm c}$ is the superconducting transition temperature and $E_{\rm SO}$ is a typical energy of the spin-orbit splitting. 
This condition is reasonable in STO since the superconducting transition temperature is extremely small, i.e., about 0.3 K. 
Then, the effect of the Rashba splitting in the FE phase is reflected to the superconductivity. 
Here, we neglect the interorbital pairing because the interorbital interaction is generally weak and does not alter qualitative results. 
Furthermore, we ignore the parity mixing of Cooper pairs in the ferroelectric (FE) phase, since the stability of FE superconductivity is hardly affected by an induced $p$-wave component. 

The impact of the applied magnetic field is included as the Zeeman coupling term, 
\begin{equation}
\mathcal{H}_{\rm Z} = - \mu_{\rm B} \sum_{{\bm k},l,\sigma, \sigma'} \vector{H} \cdot \vector{\sigma}_{\sigma \sigma'} c_{{\bm k},l\sigma}^{\dag}c_{{\bm k},l\sigma'}, 
\label{eq:H_zeeman_STO}
\end{equation}
where $\vector{\sigma}$ is the Pauli matrix and $\mu_{\rm B}$ is the Bohr magneton. 
In the superconducting STO, the superfluid density $n_s$ is small \cite{Collignon_STO}, and hence the penetration depth $\lambda_{\rm L} \propto n_{s}^{-1/2}$ is large. 
Thus, STO is a superconductor close to type-II limit with an extremely large Ginzburg-Landau parameter $\kappa_{\rm GL} \gg 1$. 
Therefore, it is justified to assume a uniform magnetic field in the bulk superconducting STO. 
It would be desirable to include the gauge interaction with the vector potential in addition to the Zeeman coupling term $\mathcal{H}_{\rm Z}$. 
The importance of the orbital depairing effect is represented by the Maki parameter $\alpha_{\rm M} \propto \Delta/\epsilon_{\rm F}$, where $\Delta$ is the superconducting gap and $\epsilon_{\rm F}$ is the Fermi energy. 
When $\alpha_{\rm M} > 1$, the orbital depairing effect is suppressed and the superconducting state is destroyed mainly due to the Pauli depairing effect. 
In the superconducting STO, $\epsilon_{\rm F}$ is extremely small and hence $\alpha_{\rm M}$ can be large. 
Thus, we assume that the orbital depairing effect is not qualitatively important in the dilute superconducting STO. 
Indeed, the upper critical field exceeding the Pauli limit in some doped STO \cite{ayino2018evidence,ahadi2019enhancing} indicates a strong impact of the Pauli depairing effect on the superconductivity. 
In the following, we fix the magnetic field in a direction parallel to the polar [001] axis, i.e., $\bm{H}=(0,0,H_z)$. 
Thus, an asymmetric deformation of the Rashba split Fermi surface, which is destructive for the FE superconductivity \cite{PhysRevB.98.024521}, is not induced under the magnetic field. 

\subsection{\label{sec:BCS}Mean-field theory for superconductivity}

We investigate the superconductivity in STO by means of the mean-field theory. 
In the following discussion, we fix $\bm{q}=\bm{0}$ since Fulde-Ferrell-Larkin-Ovchinnikov (FFLO) superconductivity with finite $\bm{q}$ is not stabilized in our model when the magnetic field is applied along the polar axis. 
The pairing interaction $\mathcal{H}_{\rm pair}$ is approximated as
\begin{align}
\mathcal{H}_{\rm pair} 
&= -\frac{V_{s}}{N} \sum_{{\bm k},{\bm k}',l}
c_{{\bm k}, l \uparrow}^{\dag} c_{-{\bm k}, l \downarrow}^{\dag}
c_{-{\bm k}', l \downarrow} c_{{\bm k}', l \uparrow}\nonumber\\
&\approx \sum_{{\bm k},l} \left(
\Delta_{l} c_{{\bm k}, l \uparrow}^{\dag} c_{-{\bm k}, l \downarrow}^{\dag}
+ \rm{H.c.} \right) 
+\frac{N}{V_s}\sum_{l}|\Delta_{l}|^2, 
\end{align}
by introducing the orbital-dependent superconducting order parameters, 
\begin{equation}
\Delta_{l} = - \frac{V_{s}}{N} \sum_{{\bm k}}
\average{c_{-{\bm k},l\downarrow} c_{{\bm k},l\uparrow}} .
\label{eq:gap_STO}
\end{equation}
To describe the total Hamiltonian $\mathcal{H} 
=\mathcal{H}_{\rm 0}+ \mathcal{H}_{\rm pol} + \mathcal{H}_{\rm Z} + \mathcal{H}_{\rm pair}$ in a matrix form, we define the vector operator as follows: 
\begin{eqnarray}
 \hat{C}_{\bm{k}}^\dag =&& ( 
c_{{\bm k},yz\uparrow}^{\dag}, c_{{\bm k},xz\uparrow}^{\dag}, 
c_{{\bm k},xy\uparrow}^{\dag}, c_{{\bm k},yz\downarrow}^{\dag}, 
c_{{\bm k},xz\downarrow}^{\dag}, c_{{\bm k},xy\downarrow}^{\dag}, c_{-{\bm k},yz\uparrow}, \nonumber\\
&& c_{-{\bm k},xz\uparrow}, 
c_{-{\bm k},xy\uparrow}, c_{-{\bm k},yz\downarrow}, 
c_{-{\bm k},xz\downarrow}, c_{-{\bm k},xy\downarrow}).
\end{eqnarray}
Then, we obtain the mean-field Hamiltonian in the matrix form
\begin{eqnarray}
\mathcal{H} =&& \frac{1}{2}\sum_{{\bm k}}
\hat{C}_{\bm k}^{\dag} \hat{\mathcal{H}}_{\rm BdG}(\vector{k}) \hat{C}_{\bm k} \nonumber\\
&&+ \frac{N}{V_s} \sum_{l} |\Delta_{l}|^2 + \sum_{\bm{k},l} \left( \varepsilon_l(\bm{k})-\mu \right) .
\end{eqnarray}
The Bogoliubov-de Gennes (BdG) Hamiltonian $\hat{\mathcal{H}}_{\rm BdG}(\vector{k})$ is described as 
\begin{equation}
 \hat{\mathcal{H}}_{\rm BdG}(\vector{k})= 
    \begin{pmatrix}
       \hat{\mathcal{H}}_{\rm N}(\vector{k}) & \hat{\Delta}  \\ 
      \hat{\Delta}^{\dag} &  -\hat{\mathcal{H}}_{\rm N}^{\rm T}(-\vector{k})
    \end{pmatrix} , 
\end{equation}
by using the normal state part 
\begin{widetext}
\begin{equation}
\hat{\mathcal{H}}_{\rm N}(\bm{k}) = 
\begin{pmatrix}
 \xi_{yz}(\bm{k})-h_z &i\lambda/2 & 
 \zeta_{yz} (\bm{k}) & 0 & 
 0 & -\lambda/2 \\ 
 -i\lambda/2 & \xi_{xz}(\bm{k})-h_z & 
 \zeta_{xz}(\bm{k}) & 0 & 
 0 & i\lambda/2 \\ 
 \zeta_{yz}^*(\bm{k}) & \zeta_{xz}^*(\bm{k}) & 
 \xi_{xy}(\bm{k})-h_z & \lambda/2 & 
 -i\lambda/2 & 0 \\
 0 & 0 & 
 \lambda/2 & \xi_{yz}(\bm{k}) + h_z & 
 -i\lambda/2 & \zeta_{yz}(\bm{k}) \\ 
 0 & 0 & 
 i\lambda/2 & i\lambda/2 & 
 \xi_{xz}(\bm{k}) + h_z & \zeta_{xz}(\bm{k}) \\ 
 -\lambda/2 & -i\lambda/2 & 
 0 & \zeta_{yz}^*(\bm{k}) & 
 \zeta_{xz}^*(\bm{k}) & \xi_{xy}(\bm{k}) + h_z
\end{pmatrix} ,
\label{eq:HN_matrix}
\end{equation}
\end{widetext}
and the pairing part
\begin{equation}
\hat{\Delta} = 
\begin{pmatrix}
 0 & 0 & 0 & \Delta_{yz} & 0 & 0 \\ 
 0 & 0 & 0 & 0 & \Delta_{xz} & 0 \\ 
 0 & 0 & 0 & 0 & 0 & \Delta_{xy} \\ 
 -\Delta_{yz} & 0 & 0 & 0 & 0 & 0 \\ 
 0 & -\Delta_{xz} & 0 & 0 & 0 & 0 \\ 
 0 & 0 & -\Delta_{xy} & 0 & 0 & 0 
\end{pmatrix} .
\label{eq:HN_matrix}
\end{equation}
Here, we abbreviate as $\xi_{l}(\bm{k})\equiv\varepsilon_{l}(\bm{k})-\mu$ and $h_z \equiv\mu_{\rm B} H_z$. 

Then, we carry out the Bogoliubov transformation
\begin{eqnarray}
c_{{\bm k},l\sigma} &=& \sum_{\nu,\tau} \left( u_{{\bm k},l\sigma}^{(\nu\tau)} \alpha_{{\bm k},\nu\tau} - v_{-{\bm k},l\sigma}^{(\nu\tau)*} \alpha_{-{\bm k},\nu\tau}^{\dag} \right) , \\
c_{-{\bm k},l\sigma}^{\dag} &=& \sum_{\nu,\tau} \left( v_{{\bm k},l\sigma}^{(\nu\tau)} \alpha_{{\bm k},\nu\tau} + u_{-{\bm k},l\sigma}^{(\nu\tau)*} \alpha_{-{\bm k},\nu\tau}^{\dag} \right) ,
\end{eqnarray}
where $\alpha_{{\bm k},\nu\tau}$ is the annihilation operator for a Bogoliubov quasiparticle with momentum $\bm{k}$, pseudoorbital $\nu=yz,xz,xy$, and pseudospin $\tau=\uparrow,\downarrow$.  
Thus, Eq. (\ref{eq:gap_STO}) is rewritten as 
\begin{equation}
\Delta_{l} = - \frac{V_{s}}{N} \sum_{{\bm k},\sigma,\nu,\tau}
\sigma_z v_{{\bm k},l\overline{\sigma}}^{(\nu\tau) *} u_{{\bm k},l\sigma}^{(\nu\tau)}f[\sigma_z E_{\nu\tau}(\bm{k})],
\label{eq:gap_bogo_STO}
\end{equation}
where $f[E]$ is the Fermi-Dirac distribution function and $E_{\nu\tau}(\bm{k})$ is the energy of a Bogoliubov quasiparticle. 
$\sigma_z=\pm1$ for $\sigma=\uparrow,\downarrow$. 
Equations (\ref{eq:gap_bogo_STO}) are the simultaneous gap equations to be solved numerically. 
In the Bogoliubov quasiparticle picture, the total Hamiltonian is described as 
\begin{eqnarray}
\mathcal{H} =&&
\sum_{{\bm k},\nu,\tau}  E_{\nu\tau}(\bm{k}) \left( \alpha_{{\bm k},\nu\tau}^{\dag} \alpha_{{\bm k},\nu\tau} - \frac{1}{2} \right) \nonumber\\
&&+ \frac{N}{V_s} \sum_{l} |\Delta_{l}|^2 + \sum_{\bm{k},l}\xi_l(\bm{k}) . 
\end{eqnarray}
Therefore, the electronic free energy per Ti site is obtained as 
\begin{eqnarray}
\mathcal{F}_{\rm ele}[\bm{\Delta},P] =&& -\frac{1}{N\beta} \sum_{{\bm k},\nu,\tau} \left[ \ln \left(1+e^{-\beta E_{\nu \tau}(\bm{k})} \right) + \frac{\beta E_{\nu \tau}(\bm{k})}{2} \right] \nonumber\\ &&+ \frac{1}{V_s} \sum_{l} |\Delta_{l}|^2 + \frac{1}{N} \sum_{\bm{k},l} \xi_l(\bm{k}) + \mu n, 
\label{eq:Helmholtz_STO}
\end{eqnarray}
where $\bm{\Delta}=(\Delta_{yz}, \Delta_{xz}, \Delta_{xy})$ and $\beta=1/T$ is the inverse temperature. 
The last term of Eq. (\ref{eq:Helmholtz_STO}) is necessary 
since the carrier density per Ti site is fixed as $n$ instead of the chemical potential $\mu$. 
Using $\bm{\Delta}$ obtained by solving Eqs. (\ref{eq:gap_bogo_STO}), we calculate the electronic part of the free energy $\mathcal{F}_{\rm ele}[\bm{\Delta},P]$ from Eq. (\ref{eq:Helmholtz_STO}).

\subsection{\label{sec:Free_pol}Polar instability}
In order to discuss the FE-like structural phase transition, we take into account the Landau free energy arising form polar lattice distortion as follows: 
\begin{equation}
\mathcal{F}_{\rm lat} [P] = \frac{1}{2} \kappa_{2} P^2 + \frac{1}{4} \kappa_{4} P^4 + \frac{1}{6} \kappa_{6} P^6, 
\end{equation}
where $\kappa_{2}$, $\kappa_{4}$, and $\kappa_{6}$ are the lattice parameters which describe the elasticity of the lattice. 
The temperature dependence of the lattice parameters is ignored, consistent with the fact that the dielectric constant is almost temperature-independent  in the quantum PE STO \cite{PhysRevB.19.3593}. 

The total free energy including the contributions of both electrons and lattice is given by
\begin{equation}
\mathcal{F}[\bm{\Delta},P] = \mathcal{F}_{\rm ele}[\bm{\Delta},P] + \mathcal{F}_{\rm lat}[P] . 
\end{equation}
The thermodynamically stable state is determined by minimizing the free energy 
$\mathcal{F}[\bm{\Delta},P]$ with respect to $\bm{\Delta}$ and $P$. 
The FE superconducting state is realized when both $\bm{\Delta}$ and $P$ have finite values. 
The values of the phenomenological lattice parameters $\kappa_2$, $\kappa_4$, and $\kappa_6$ are determined as follows. 
The lattice parameters $\kappa_{4}$ and $\kappa_{6}$ are introduced to cut off the FE order parameter $\gamma \propto P$ in a realistic regime. 
In this study, we set $\kappa_{4}$ and $\kappa_{6}$ so as to satisfy $\gamma/t_1\lesssim 0.20$ in agreement with the first-principles estimation of $\gamma$ for the SrTiO$_3$/LaAlO$_3$ interface \cite{PhysRevB.87.161102}. 
The choice of $\kappa_4$ and $\kappa_6$ hardly alters the results of this paper. 
The value of $\kappa_{2}$ is determined so as to realize a PE normal state near a FE phase transition point. 
This condition expresses a situation of the PE STO which is moved toward the vicinity of the FE quantum critical point, for example, by Ca substitution \cite{PhysRevLett.52.2289} or isotopic substitution of O\cite{PhysRevLett.82.3540}. 
Then, we investigate the feasibility of a FE transition caused by the superconductivity.  

The phenomenological parameters $\kappa_2$, $\kappa_4$, and $\kappa_6$ might be derived from the microscopic Hamiltonian for the optical phonon excitations coupled to the conduction electrons, in which the FE transition should be driven by the dipolar interaction \cite{cowley1980structural,PhysRevMaterials.2.104804}. 
They are phenomenologically introduced in this study, and more microscopic study for the ferroelectricity in the superconducting STO is left for a future work.

\section{\label{sec:Multiorbital}Multiorbital and multiband effect}
Before showing results for the FE superconductivity, we here clarify effects of the multiorbital and multiband electronic structure in STO. 

\subsection{\label{sec:ASOC}Unconventional Rashba spin-orbit coupling}

\begin{figure}[htbp]
   \centering
   \includegraphics[width=87mm,clip]{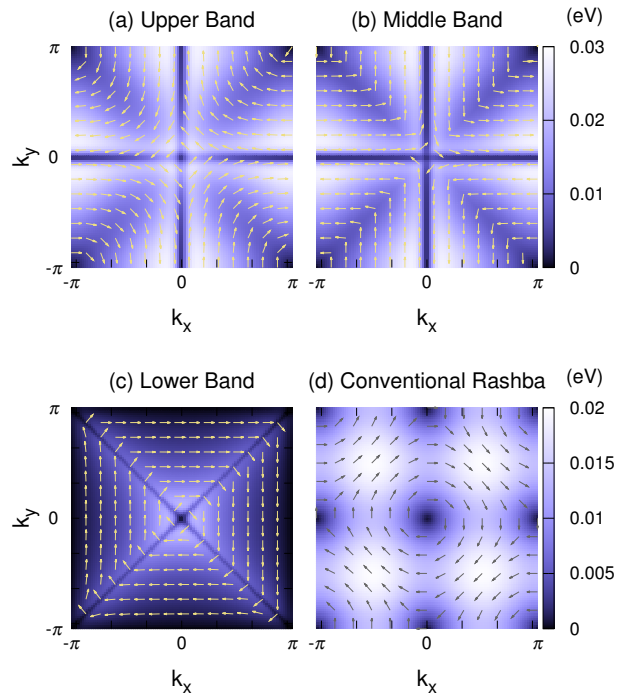}
\caption{The magnitude of the spin-orbit splitting in the (a) upper band, (b) middle band, and (c) lower band at $k_z=0$ with $\gamma/t_1=0.105$. The arrows show the direction of the $g$-vector. (d) $\bm{k}$ dependence of the conventional Rashba ASOC given by Eq. (\ref{eq:g_conventional}) with $\gamma/t_1=0.105$. 
\label{fig:STO_Rashba} }
\end{figure}

First, we investigate the multiorbital effect on the Rashba spin-orbit splitting in the FE STO. 
We elucidate the nature of the Rashba ASOC by calculating the energy spectrum in the normal state $\mathcal{E}_{m}(\bm{k})$ [$\mathcal{E}_{m}(\bm{k})\leq\mathcal{E}_{m'}(\bm{k})$ for $m<m'$] and wave functions. 
In the presence of the inversion symmetry ($\gamma=0$), the two-fold degeneracy holds as $\mathcal{E}_{1}(\bm{k})=\mathcal{E}_{2}(\bm{k})$, $\mathcal{E}_{3}(\bm{k})=\mathcal{E}_{4}(\bm{k})$, and $\mathcal{E}_{5}(\bm{k})=\mathcal{E}_{6}(\bm{k})$. 
On the other hand, Rashba-type spin-orbit splitting is induced by the polar inversion symmetry breaking ($\gamma\neq0$) as $\mathcal{E}_{1}(\bm{k})<\mathcal{E}_{2}(\bm{k})$, $\mathcal{E}_{3}(\bm{k})<\mathcal{E}_{4}(\bm{k})$, and $\mathcal{E}_{5}(\bm{k})<\mathcal{E}_{6}(\bm{k})$ except for at the time-reversal invariant momentum. 
Spin direction of each Rashba split bands is calculated by taking the average  $\bm{S}_{\alpha}(\bm{k})=\average{\sum_{l}\sum_{\sigma,\sigma'}\bm{\sigma}_{\sigma\sigma'}c_{\bm{k},l\sigma}^{\dag}c_{\bm{k},l\sigma'}}_{\alpha}$ for the wave function of the $\alpha$-th band. 
Figures \ref{fig:STO_Rashba}(a), \ref{fig:STO_Rashba}(b) and \ref{fig:STO_Rashba}(c) show the magnitude of the spin-orbit splitting $\delta\mathcal{E}_{\alpha}(\bm{k})=\mathcal{E}_{2\alpha}(\bm{k})-\mathcal{E}_{2\alpha-1}(\bm{k})$ ($\alpha=3,2,1$) and the direction of the $g$-vector defined as $\bm{g}_{\alpha}(\bm{k})=\delta\mathcal{E}_{\alpha}(\bm{k})\bm{S}_{\alpha}(\bm{k})$ for each Rashba split bands. 
Note that the upper, middle, and lower bands are denoted by $\alpha=3$, 2, and 1, respectively. 
We see that the $\bm{k}$-dependence of the Rashba spin-orbit splitting in STO is significantly different from that of the conventional Rashba ASOC with $\bm{g}(\bm{k})=\left(\sin k_y, -\sin k_x, 0 \right)$ [Fig. \ref{fig:STO_Rashba}(d)]. 
The spin-orbit splitting in the lower band is large at $\bm{k}$ slightly away from the $\Gamma$-M line, whereas that in the middle or upper band is large at $\bm{k}$ slightly away from the $\Gamma$-X line. 
In particular, the spin-orbit splitting in the lower band is maximized near the $\Gamma$-point where the spin-orbit splitting of the conventional Rashba ASOC is tiny. 
Moreover, the $g$-vectors of the lower and middle bands are almost parallel to the [100] or [010] axis, and rapidly rotates by $\pi/2$ at the $\Gamma$-M line. 
The origin of this unconventional Rashba ASOC is explained by combined analysis of the perturbation expansion for the LS coupling and the basis transformation to the total angular momentum space [see Appendix]. 
The unconventional Rashba ASOC characteristic of the multiorbital system gives impacts on superconductivity, as we show below. 

\subsection{\label{sec:Hc2}Enhanced upper critical field of dilute superconductivity}

\begin{figure}[htbp]
   \centering
   \includegraphics[width=90mm,clip]{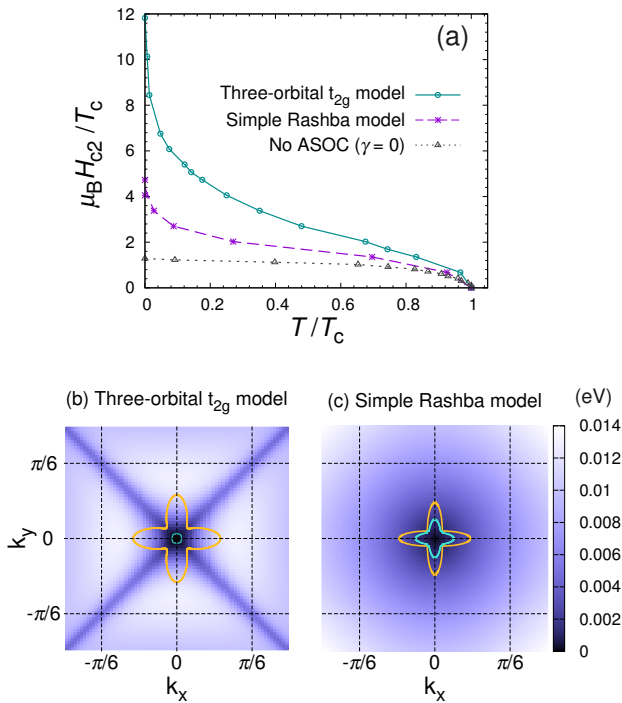}
\caption{(a) The temperature dependence of the upper critical field $\mu_{\rm B}H_{c2}$. The magnetic filed is applied along the [001] axis. The green solid line (purple dashed line) shows the upper critical field calculated for the three-orbital model (simple Rashba model). The gray dotted line shows the upper critical field of the three-orbital model with $\gamma=0$. 
(b) and (c) Illustration of the Fermi surface and the magnitude of the spin-orbit splitting of the lower band for the (b) three-orbital model and (c) simple Rashba model. The carrier density and odd-parity hopping integral are set to $n=5.0\times10^{-5}$ and $\gamma/t_1=0.105$, respectively. 
\label{fig:Hc2} }
\end{figure}

Next, we discuss an impact of the multiorbital electronic structure on the dilute superconductivity in STO. 
As shown in Sec. \ref{sec:ASOC}, unconventional Rashba ASOC is induced by ferroelectricity as a consequence of the multiorbital effect. 
Unlike the conventional Rashba effect, the Rashba splitting in the lower band is maximized near the $\Gamma$-point [Fig. \ref{fig:STO_Rashba}(c)]. 
On the other hand, the Pauli depairing effect of a Rashba superconductor is suppressed under a magnetic field parallel to the polar axis \cite{EdelsteinJETP,PhysRevLett.75.2004,PhysRevLett.92.097001,fujimoto2007electron,fujimoto2007fermi}. 
This is because the BCS-type Cooper pairing is possible even under the magnetic field thanks to the Rashba-type spin-momentum locking. 
Thus, it is implied that the Pauli depairing effect in a dilute superconducting state with a tiny Fermi surface should be strongly suppressed compared to the case of a conventional Rashba superconductor. 

To test the above expectation, we introduce a simple Hamiltonian with conventional Rashba ASOC as follows: 
\begin{align} 
\tilde{\mathcal{H}} =& \tilde{\mathcal{H}}_{0} + \mathcal{H}_{\rm Z}+ \mathcal{H}_{\rm pair} , \label{eq:H_eff_Rashba} \\
\tilde{\mathcal{H}}_{0} =& \sum_{{\bm k},l,\sigma} \left( \varepsilon_l({\bm k}) - \mu \right)  c_{{\bm k}, l \sigma}^{\dag}c_{{\bm k}, l \sigma} \nonumber\\
&+ \alpha \sum_{\bm{k},l,\sigma,\sigma'} \bm{g}(\bm{k}) \cdot \bm{\sigma}_{\sigma\sigma'} c_{\bm{k},l\sigma}^{\dag} c_{\bm{k},l\sigma'} \label{eq:g_conventional},
\end{align}
where the $g$-vector is assumed to be $\bm{g}(\bm{k})=\left( \sin k_y, -\sin k_x, 0 \right)$ and diagonal in the orbital space. 
Based on the perturbation analysis for the LS coupling [see Eq. (\ref{eq:STO_g1_LS}) in Appendix], we assume $\alpha=2\lambda\gamma/t_1$ in the following discussion. 
Here, we compare this model with our three-orbital model for STO to illuminate the multiorbital effect which is appropriately taken into account in the latter. 
Figure \ref{fig:Hc2}(a) shows the temperature dependence of the upper critical field $\mu_{\rm B}H_{c2}$ in the dilute single-band regime ($n=5.0\times10^{-5}$) where the Fermi surface is only composed of the lower band. 
Since the lattice constant of STO $\sim$ 3.905 {\AA} is chosen as the unit of length, $n=5.0 \times 10^{-5}$ corresponds to $8.40\times 10^{17}$ cm$^{-3}$.  
The paring interaction is chosen to $V_s/t_1=0.28$, hence the superconducting transition temperature is set to $T_{\rm c}=0.00098t_1\sim3.0$ K at $\gamma=0$ and $H_z=0$. 
When we adopt $\gamma=0.105 t_1$, a typical energy of spin-orbit splitting is $E_{\rm SO}\sim\lambda\gamma/t_1\sim0.01$.
Then, $T_{\rm c}\ll E_{\rm SO}$ is satisfied and the effect of the Rashba splitting is reflected to the superconductivity.
Note that the superconducting transition temperature is set to be larger than the realistic value $T_{\rm c}\sim0.3$ K of doped STO to reduce the cost of numerical calculation. 
In both models, the upper critical field of the noncentrosymmetric superconductivity with Rashba splitting (green solid line and purple dashed line) is enhanced compared to that of the centrosymmetric superconductivity without ASOC (gray dotted line). 
Furthermore, we see that the upper critical field of the three-orbital model for STO [Eqs. (\ref{eq:H_t2g_STO}) and (\ref{eq:H_pol})] is larger than that of the simple Rashba model [Eq. (\ref{eq:H_eff_Rashba})]. 
The origin of this enhanced upper critical field can be attributed to the Fermi surface anisotropy and large spin-orbit splitting. 
As shown in Figs. \ref{fig:Hc2}(b) and \ref{fig:Hc2}(c), the Fermi surfaces of the lower band show $d_{x^2-y^2}$-wave like anisotropy. 
The unconventional Rashba ASOC in the three-orbital model induces a large spin-orbit splitting particularly in the region near $\bm{k}\parallel[100]$. 
The DOS at the Fermi energy mainly comes from this region. 
Furthermore, the magnitude of the spin-orbit splitting is maximized near the $\Gamma$-point, where the Fermi surfaces in the dilute region are located [Fig. \ref{fig:Hc2}(b)]. 
Therefore, the upper critical field is drastically enhanced thanks to the strong spin-momentum locking on the Fermi surface. 
In contrast, the conventional Rashba ASOC induces a small spin-orbit splitting around the $\Gamma$-point as shown in Fig. \ref{fig:Hc2}(c). 
Thus, the enhancement of the upper critical field by spin-orbit splitting is less pronounced than that in the three-orbital model. 

\subsection{\label{sec:Lifshitz}Lifshitz transitions and superconductivity}

\begin{figure}[htbp]
   \centering
   \includegraphics[width=68mm,clip]{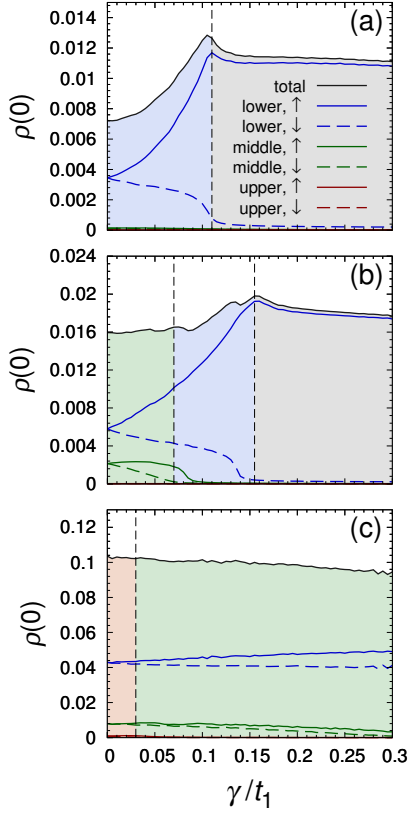}
\caption{DOS at the Fermi energy $\rho(0)$ as a function of the odd-parity hopping integral $\gamma$. The partial DOS for the Rashba split $t_{2g}$ bands are also shown. The black dashed vertical lines represent the Lifshitz transition point. Different colored regions indicate different phases which are distinguished by the Lifshitz transitions. The colors correspond to the background colors in Fig. \ref{fig:Band}. The carrier density is set to (a) $n=5.0\times10^{-5}$, (b) $n=2.0\times10^{-4}$, and (c) $n=1.0\times10^{-2}$, respectively. 
\label{fig:Lifshitz} }
\end{figure}

In this section, we discuss the ferroelectricity-induced Lifshitz transition and its effect on superconductivity. 
Upon decreasing the carrier density in the FE phase, the Fermi energy becomes lower than the crossing point of the spin-orbit split bands at the $\Gamma$-point [see Fig. \ref{fig:Band}], and thus the topology of Fermi surfaces is changed in stages.  
These Lifshitz transitions enhance the DOS due to an effective reduction of dimensionality \cite{PhysRevLett.98.167002}, and leads to the stabilization of a FE superconducting state \cite{PhysRevB.98.024521}. 

\begin{figure}[htbp]
   \centering
   \includegraphics[width=78mm,clip]{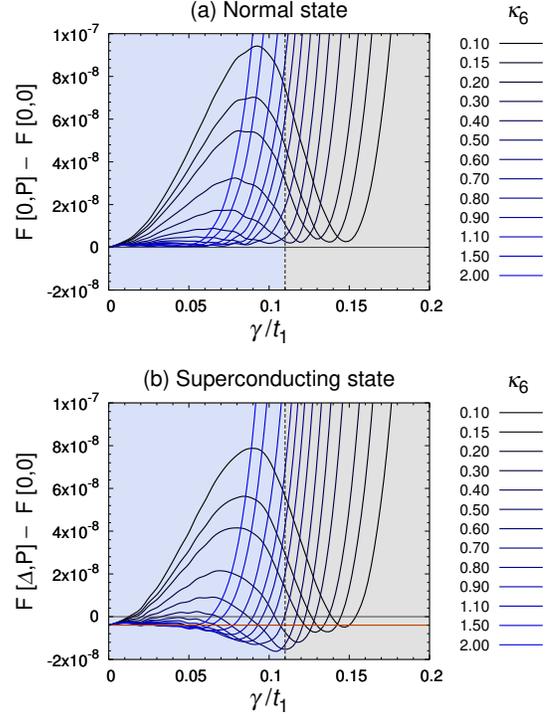}
\caption{The free energy as a function of the odd-parity hopping integral $\gamma$ for several values of the cutoff lattice parameter $\kappa_{6}$. The carrier density is set to $n=5.0\times10^{-5}$, i.e., a single-band regime. 
The other lattice parameters are chosen as $(\kappa_2, \kappa_4) = (6.75\times10^{-5}, 0)$. The temperature and magnetic field are set to $T=1.0\times10^{-10}$ and $\bm{H}=0$, respectively. The black dashed vertical line shows the Lifshitz transition point. (a) $\gamma\propto P$ dependence of $\delta\mathcal{F}[0,P]=\mathcal{F}[0,P]-\mathcal{F}[0,0]$. Since $\delta\mathcal{F}[0,P]>0$ is satisfied in the whole range of $\kappa_{6}$, the PE normal state is realized. (b) $\gamma\propto P$ dependence of $\delta\mathcal{F}[\bm{\Delta},P]=\mathcal{F}[\bm{\Delta},P]-\mathcal{F}[0,0]$. The stabilization condition of the FE superconducting state, i.e., $\delta\mathcal{F}[\bm{\Delta},P]<\delta\mathcal{F}[\bm{\Delta},0]<0$, is satisfied under the red horizontal line. 
\label{fig:FESC_dilute} }
\end{figure}

Figure \ref{fig:Lifshitz} shows the DOS at the Fermi energy $\rho(0)$ as a function of $\gamma$. 
In the single-band regime ($n=$5.0$\times$10$^{-5}$), $\rho(0)$ is maximized at the Lifshitz transition point of the lowest band $\gamma=\gamma_{c1}$ [Fig. \ref{fig:Lifshitz}(a)], thanks to the effective reduction of the dimensionality. 
Consequently, the superconductivity is enhanced at the Lifshitz transition point $\gamma_{c1}$. 
Figure \ref{fig:FESC_dilute}(b) shows the $\gamma\propto P$ dependence of the FE superconducting condensation energy $\delta\mathcal{F}[\bm{\Delta},P]=\mathcal{F}[\bm{\Delta},P]-\mathcal{F}[0,0]$ for various values of the cutoff lattice parameter $\kappa_{6}$. 
Here, $\bm{\Delta}$  is optimized under fixed $P$. 
We see that the stabilization condition of the FE superconducting sate, i.e., $\delta\mathcal{F}[\bm{\Delta},P]<\delta\mathcal{F}[\bm{\Delta},0]<0$, is satisfied in a wide range of lattice parameters, although the normal state is PE [see Fig. \ref{fig:FESC_dilute}(a)]. 
This means that the FE superconductivity is stable at zero magnetic field in the dilute region. 

Next, we discuss the effects of Lifshitz transitions of the middle and upper bands. 
When the Fermi energy in the PE phase is slightly higher than the bottom of the bands, the Lifshitz transition is induced by the  ferroelectricity. 
However, we see that these Lifshitz transitions are not significantly reflected in the total DOS $\rho(0)$ [Figs. \ref{fig:Lifshitz}(b) and \ref{fig:Lifshitz}(c)]. 
Although the partial DOS for the middle or upper band is enhanced as approaching to the Lifshitz transitions, the contribution of the partial DOS is very small compared to that of the lower band. 
Therefore, the total DOS $\rho(0)$ is not drastically enhanced by these Lifshitz transitions, and hence the FE superconducting state is hardly stabilized at zero magnetic field in relatively high carrier density two- or three-band regimes. 
The phase diagram of the FE superconductivity depends on what band causes the Lifshitz transition.


\section{\label{sec:FESC}Ferroelectric Superconductivity}
\subsection{\label{sec:Phase}Phase diagram}

\begin{figure*}[htbp]
   \centering
   \includegraphics[width=180mm,clip]{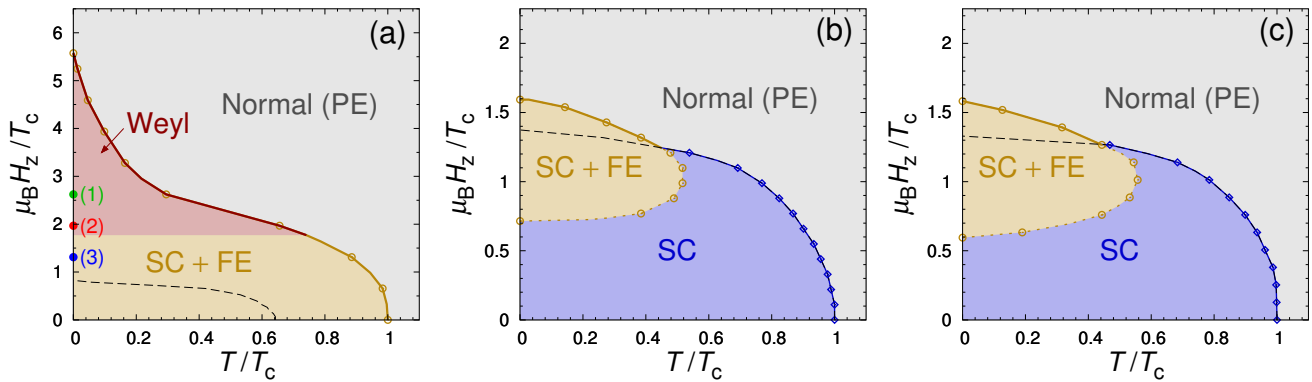}
\caption{Phase diagrams in a (a) single-band regime ($n=5.0\times10^{-5}$), (b) two-band regime ($n=3.2\times10^{-3}$), and (c) three-band regime ($n=1.8\times10^{-2}$). 
Band structures corresponding to these carrier densities are shown in Fig. \ref{fig:FS_ASOC}. 
The yellow or red solid (dotted) line shows the first-order (second-order) FE phase transition line. 
The black dashed line indicates the PE superconducting transition line obtained by assuming $\gamma=0$. The red colored region in (a) illustrates the Weyl superconducting phase. The pairing interaction and the lattice parameters are set to (a) ($V_{\rm s}/t_1, \kappa_2, \kappa_4, \kappa_6) = (0.28, 6.75\times10^{-5}, 0, 0.50)$ (b) ($V_{\rm s}/t_1, \kappa_2, \kappa_4, \kappa_6) = (0.51, 1.00\times10^{-2}, 0, 0)$ and (c) ($V_{\rm s}/t_1, \kappa_2, \kappa_4, \kappa_6) = (0.77, 5.30\times10^{-2}, 0, 0)$ respectively. 
The temperature $T$ and the magnetic field $\mu_{\rm B}H_z$ are normalized by the superconducting transition temperature $T_{\rm c}$ at zero magnetic field. 
\label{fig:Phase} }
\end{figure*}

\begin{figure}[bp]
\centering
   \includegraphics[width=86mm,clip]{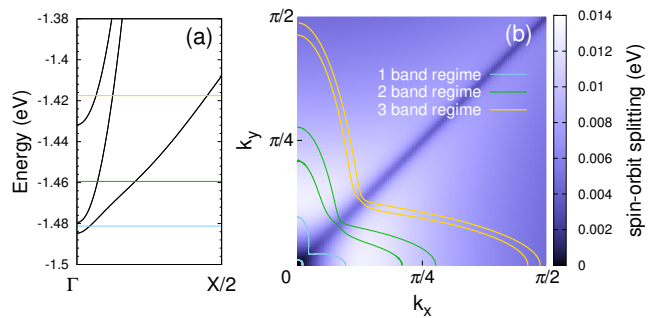}
   \caption{(a) Band structure for $\gamma=0$. The blue, green and yellow horizontal lines illustrate the Fermi energy in a single-band regime ($n=5.0\times10^{-5}$), two-band regime ($n=3.2\times10^{-3}$), and three-band regime ($n=1.8\times10^{-2}$), respectively. 
(b) Illustration of the Rashba split Fermi surfaces of the lower band ($\gamma=29.1$ meV) in the single-, two-, and three-band regime, overwritten on the magnitude of spin-orbit splitting. 
Color of Fermi surfaces corresponds to the colored lines in (a). 
\label{fig:FS_ASOC} }
\end{figure}

Figure \ref{fig:Phase} shows the magnetic field versus temperature phase diagrams in three different carrier density regimes which are distinguished by the number of Fermi surfaces [see Fig. \ref{fig:FS_ASOC}(a)]. 
In the single-band regime, the FE superconducting state is stabilized at zero magnetic field [Fig. \ref{fig:Phase}(a)]. 
This is a consequence of a Lifshitz transition induced by the ferroelectricity as shown in Sec. \ref{sec:Lifshitz}.  
On the other hand, the zero field FE superconductivity is not realized in the two- or three-band regime due to the small contribution of the middle or upper band to the total DOS. 
Because of the multiband nature of STO, the zero field FE superconductivity is possible only in the dilute region where the Lifhisitz transition of the lower band can be induced by the ferroelectricity. 

Irrespective of the carrier density, the FE superconducting state is stabilized under the magnetic field, despite an absence of the zero field FE superconducting phase in the two- or three-band regime [Figs. \ref{fig:Phase}(b) and \ref{fig:Phase}(c)]. 
This magnetic-field-induced FE superconductivity originates from the anomalous Pauli depairing effect in noncentrosymmetric superconductors \cite{EdelsteinJETP,PhysRevLett.75.2004,PhysRevLett.92.097001,fujimoto2007electron,fujimoto2007fermi}. 
To avoid the Pauli depairing effect, superconductivity induces the FE order giving rise to the Rashba ASOC. 
Then, the upper critical field is enhanced compared to the PE state.

In particular, the enhancement of the upper critical field is remarkable in the single-band regime [Fig. \ref{fig:Phase}(a)]; 
$\mu_{\rm B} H/T_c \sim 5.5$ far exceeds the Pauli limit $\sim 1.25$. 
This is owing to the Lifshitz transition and the multiorbital effect discussed in Sec. \ref{sec:Hc2}. 
For low carrier densities, the free energy is minimized at a large value of $\gamma$ so that the Lifshitz transition of the lower band occurs [Fig. \ref{fig:FESC_dilute}(b)]. 
Consequently, the first-order FE transition occurs at the same time as the superconducting transition. 
Furthermore, the Rashba spin-orbit splitting of the lower band particularly becomes large around the $\Gamma$-point thanks to the multiorbital effect [see Figs. \ref{fig:STO_Rashba}(c) and  \ref{fig:FS_ASOC}(b)]. 
Therefore, the Pauli depairing effect is strongly suppressed and the upper critical field is strongly enhanced as shown in Sec. \ref{sec:Hc2}. 
On the other hand, the $\gamma$ and resulting spin-orbit splitting are small in a higher carrier density regime, and thus the upper critical field is not significantly enhanced [Figs. \ref{fig:Phase}(b) and \ref{fig:Phase}(c)].

\subsection{\label{sec:Weyl}Weyl superconductivity}
\begin{figure}[tbp]
\centering
   \includegraphics[width=80mm,clip]{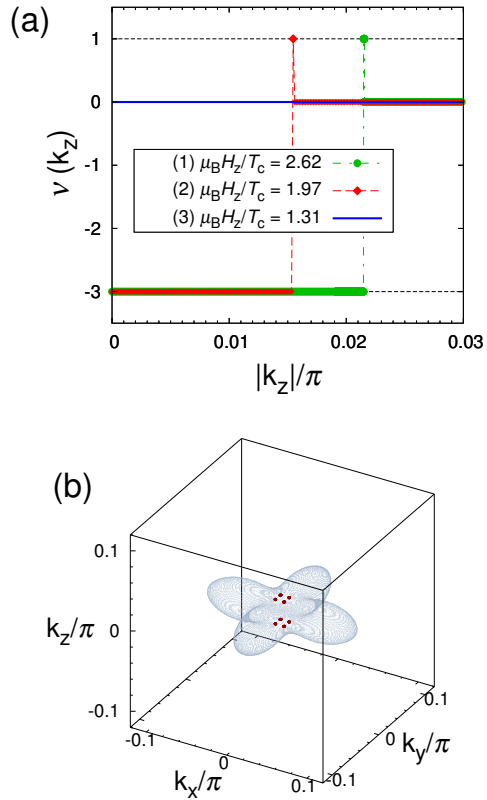}
   \caption{(a) $k_z$-dependent Chern number calculated at three magnetic fields (1), (2), and (3) in Fig. \ref{fig:Phase}(a). In the Weyl superconducting phase, $\nu(k_z)$ changes as $-3 \rightarrow 1 \rightarrow 0$. (b) Weyl nodes (red points) on the Fermi surface.  Parameters correspond to the point (1) in Fig. \ref{fig:Phase}(a). 
\label{fig:Chern} }
\end{figure}

As a consequence of the drastically enhanced upper critical field, the dilute superconducting state in STO may realize a topological Weyl superconductor. 
In a two-dimensional Rashba superconductor, a gapped topological superconducting state in class $D$ can be realized under a perpendicular magnetic field \cite{PhysRevLett.103.020401,PhysRevB.82.134521}. 
In our three dimensional case, a Weyl superconducting state, which is characterized by topologically-protected Weyl nodes, is realized in the FE phase for a wide range of the magnetic field along the polar axis. 
We identify Weyl nodes by calculating $k_z$-dependent Chern number, 
\begin{equation}
\nu(k_z)=\frac{1}{2\pi} \int dk_x dk_y F_{z}(\bm{k}), 
\end{equation}
on a two-dimensional $k_x$-$k_y$ plane \cite{PhysRevLett.49.405,kohmoto1985topological,fukui2005chern}. 
The Berry flux $F_{a}(\bm{k})$ is defined as 
\begin{equation}
F_{a}(\bm{k})=-i\epsilon^{abc} \displaystyle \sum_{E_{m}(\bm{k})<0} \partial_{k_{b}} \braket{u_{m}(\bm{k})|\partial_{k_c} u_{m}(\bm{k})}, 
\end{equation}
where the wave function of a Bogoliubov quasiparticle with energy $E_{m}(\bm{k})$ is denoted as $\ket{u_{m}(\bm{k})}$. 
Since a jump in $\nu(k_z)$ is equivalent to the sum of Weyl charges at $k_z$, we can detect Weyl charges by counting point nodes and comparing it with the jump. 
As shown in Fig. \ref{fig:Chern}(a), the Chern number jumps by $+1$ and $-4$. 
Thus, we identify five pairs of Weyl nodes and we illustrate the distribution of Weyl charges in the momentum space in Fig. \ref{fig:Chern}(b). 
One of them is located at poles of the Fermi surface. 
The rest of Weyl nodes, which are protected by $C_4$ symmetry, surrounds the above Weyl nodes with opposite Weyl charges. 
These four pairs of Weyl nodes arise as a consequence of the anisotropic Fermi surfaces due to the multiorbital effect. 
Therefore, a Weyl superconducting state with Chern number $\nu(k_z)=(+1)\times1+(-1)\times4=-3$ is obtained. 
Thus, multiorbital nature of STO leads to topological property distinct from the single-orbital topological Rashba superconductor with Chern number $\nu=\pm1$ \cite{PhysRevLett.103.020401,PhysRevB.82.134521}. 
It gives rise to three Majorana arcs in the surface state, and the zero-field thermal conductivity $\kappa_{xy}'\sim T\int dk_z\nu(k_z)$ \cite{sumiyoshi2013quantum} in STO should be larger than that in the single-orbital Rashba model. 

\section{\label{sec:Summary}Summary and conclusion}

In summary, we have studied an interplay of FE order and superconductivity in STO. 
In particular, we have proposed that the FE superconductivity is realized in STO near a FE transition point. 
The superconductivity triggers the coexisting FE order.
A key ingredient is the Rashba ASOC in the FE phase. 
By analyzing the realistic three-orbital model, we showed that the zero field FE superconductivity is stabilized only in the dilute regime where the Lifshitz transition of the lower band can be induced by the ferroelectricity.  
This result is consistent with the experimental observation in Sr$_{1-x}$Ca$_{x}$TiO$_{3-\delta}$ \cite{NatPhys.13.643-648}, which indicates the FE superconducting phase only in a dilute carrier density regime. 
Furthermore, we revealed that the FE superconducting state is stabilized under a magnetic field independent of the number of Fermi surfaces. 
This magnetic-field-induced phase appears because of the suppression of the Pauli depairing effect thanks to the Rashba-type spin-momentum locking. 
Consequently, the upper critical field is enhanced by the FE transition.  
The upper critical field is particularly large in the dilute carrier density regime, because the multiorbital effect leads to a large spin-orbit splitting distinct from the conventional Rashba model. 
Furthermore, the high magnetic field region of dilute superconducting STO is identified as a Weyl superconducting state. 
This topological phase transition is realized as a result of the multiorbital effect and Lifshitz transition, in sharp contrast to the two-dimensional single-orbital model where the FE topological superconductivity is unstable \cite{PhysRevB.98.024521}. 
These results are based on a simple BCS-type pairing interaction. 
Although we think that inclusion of dynamical electron-phonon couplings and Coulomb interactions will not dramatically alter the results, such calculation is desired and left for a future study. 

The results of this paper suggest a tunable crystal symmetry through superconductivity, in the presence of a coupling between spin, orbital, and lattice degrees of freedom. 
Most of the novel interplay of  FE-like polar inversion symmetry breaking and superconductivity was uncovered in the dilute carrier density region. 
The dilute superconductivity in STO provides an ideal platform for the FE superconductivity.

\begin{acknowledgments}
The authors are grateful to C. W. Rischau, K. Behnia, and S. Stemmer for providing their recent data and helpful discussions. 
This work was supported by Grant-in Aid for Scientific
Research on Innovative Areas “J-Physics” (JP15H05884) and
“TopologicalMaterials Science” (JP18H04225) from Japan Society for the Promotion of Science (JSPS), and by JSPS Core to Core program “Oxide Superspin (OSS)” international networking, and by JSPS KAKENHI (Grants No. JP15H05745, No. JP15K05164, No. JP18H01178, and No. JP18H05227). 
S. K. is supported by a JSPS research fellowship and by JSPS KAKENHI (Grant No. 19J22122). 
\end{acknowledgments}

\appendix*
\section{\label{sec:Appendix}Effective Rashba spin-orbit coupling}
In the Appendix, we clarify the origin of the unconventional Rashba ASOC in STO by deriving the effective Hamiltonian from two approaches. 

\subsection{\label{sec:LS}Perturbation analysis for LS coupling}
\begin{figure*}[tbp]
   \centering
   \includegraphics[width=180mm,clip]{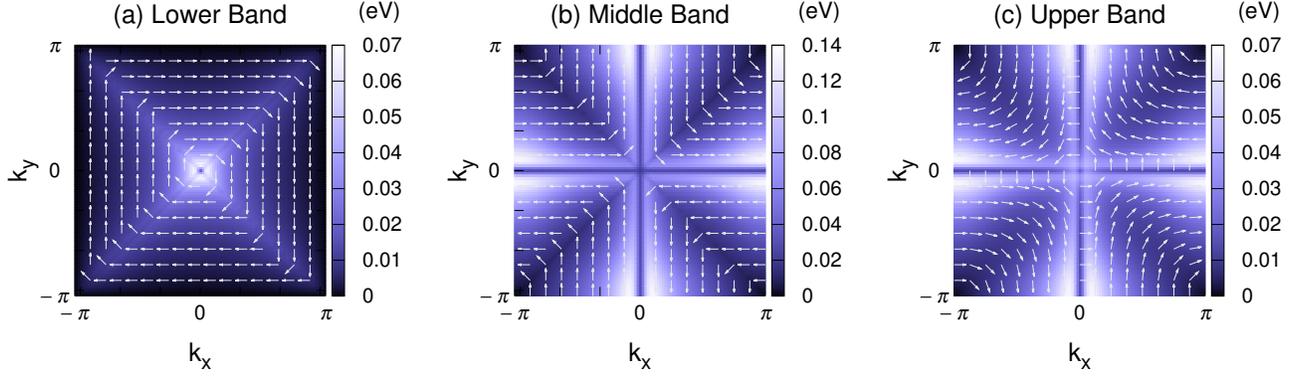}
\caption{The magnitude of the spin-orbit splitting, which is derived by the perturbation analysis for $\lambda$ and $\gamma$, in the (a) lower band, (b) middle band, and (c) upper band at $k_z=0$. The odd-parity hopping integral is set to $\gamma/t_1=0.105$. The arrows show the direction of the effective $g$-vector $\bm{g}_{\alpha}^{L}(\bm{k})$. 
\label{fig:ASOC_L} }
\end{figure*}
First, we carry out the perturbation analysis for the LS coupling $\lambda$ and the odd-parity hybridization $\gamma$. 
This analysis is valid when $\lambda$ is much smaller than other energy scales such as the band-width.  
As a result of the first-order perturbation expansion for $\lambda$, we obtain a new $\bm{k}$-dependent basis as follows: 
\begin{eqnarray}
\ket{d_{yz},\sigma}_{\bm{k},\lambda}=&& \ket{d_{yz},\sigma}  \nonumber\\
&&+ \frac{\lambda\sigma_z}{2} \left( \frac{\ket{d_{xy},\overline{\sigma}}}{\delta_{xy,yz}(\bm{k})} + \frac{i\ket{d_{xz},\sigma} }{\delta_{xz,yz}(\bm{k})} \right), \\
\ket{d_{xz},\sigma}_{\bm{k},\lambda}=&& \ket{d_{zx},\sigma} \nonumber\\
&&+ \frac{\lambda}{2} \left( \frac{i\ket{d_{xy},\overline{\sigma}}}{\delta_{xy,xz}(\bm{k})}+\frac{i\sigma_z \ket{d_{yz},\sigma}}{\delta_{xz,yz}(\bm{k})} \right), \\
\ket{d_{xy},\sigma}_{\bm{k},\lambda}=&& \ket{d_{xy},\sigma} \nonumber\\
&&+ \frac{\lambda}{2} \left( \frac{\sigma_z \ket{d_{yz},\overline{\sigma}}}{\delta_{xy,yz}(\bm{k})}+ \frac{i\ket{d_{xz},\overline{\sigma}}}{\delta_{xy,xz}(\bm{k})}  \right), 
\end{eqnarray}
where $\ket{t_{2g},\sigma}$ ($t_{2g}=d_{yz},d_{xz},d_{xy}$, $\sigma=\uparrow,\downarrow$) is the wave function of the local $t_{2g}$ orbitals and $\delta_{l,l'}(\bm{k})\equiv\varepsilon_{l}(\bm{k})-\varepsilon_{l'}(\bm{k})$. 
Then, we carry out the $\bm{k}$-dependent basis transformation for $\mathcal{H}_{0}+\mathcal{H}_{\rm pol}$ from the local $t_{2g}$ orbital space $\ket{t_{2g},\sigma}$ to the renormalized $t_{2g}$ orbital space $\ket{t_{2g},\sigma}_{\bm{k},\lambda}$. 
In addition, we perform a block diagonalization for up and down pseudospin sectors to derive the effective ASOC. 
Finally, we neglect the interorbital component since the orbital hybridizations by $\lambda$ and $\gamma$ are assumed to be small. 
Thus, in the case of a weak spin-orbit coupling, the effective Hamiltonian is described as 
\begin{align}
\tilde{\mathcal{H}}_{0}^{L} =& 
\sum_{\bm{k},\alpha,\sigma} 
\left(\varepsilon_{\alpha}^{L}({\bm k})-\mu\right) c_{{\bm k},\alpha\sigma}^{\dag} c_{{\bm k},\alpha\sigma} \nonumber\\
&+ \sum_{\bm{k},\alpha,\sigma,\sigma'} 
\bm{g}_{\alpha}^{L}(\bm{k}) \cdot \bm{\sigma}_{\sigma\sigma'} c_{{\bm k},\alpha\sigma}^{\dag} c_{{\bm k},\alpha\sigma'}, 
\label{eq:H_eff_lambda}
\end{align}
where the lower, middle, and upper bands are denoted by the band index $\alpha=1,2,3$. 
The renormalized single-particle energy $\varepsilon_{\alpha}^{L}({\bm k})$ are described as
\begin{align}
 \varepsilon_{1}^{L}(\bm{k}) &= \frac{1}{2}\Bigl( \varepsilon_{yz}(\bm{k})+\varepsilon_{xz}(\bm{k})-|\delta_{xz,yz}(\bm{k})| \Bigr), \\
\varepsilon_{2}^{L}(\bm{k}) &= \frac{1}{2} \Bigl( \varepsilon_{yz}(\bm{k})+\varepsilon_{xz}(\bm{k})+|\delta_{xz,yz}(\bm{k})| \Bigr), \\
\varepsilon_{3}^{L}(\bm{k}) &= \varepsilon_{xy}(\bm{k}), 
\end{align}
and the effective $g$-vector $\bm{g}_{\alpha}^{L}(\bm{k})$ are obtained as 
\begin{align}
\bm{g}_{1}^{L}(\bm{k})&= 2\lambda\gamma 
\begin{pmatrix}
 \sin k_y \left(\cfrac{1-\mathrm{sgn}[\delta_{xz,yz}(\bm{k})]}{\delta_{xy,xz}(\bm{k})} \right) \\
 -\sin k_x \left(\cfrac{1+\mathrm{sgn}[\delta_{xz,yz}(\bm{k})]}{\delta_{xy,yz}(\bm{k})} \right) \\
 0
\end{pmatrix} , \label{eq:STO_g1_LS} \\
\bm{g}_{2}^{L}(\bm{k})&= 2\lambda\gamma 
\begin{pmatrix}
 \sin k_y \left(\cfrac{1+\mathrm{sgn}[\delta_{xz,yz}(\bm{k})]}{\delta_{xy,xz}(\bm{k})} \right) \\
 -\sin k_x \left(\cfrac{1-\mathrm{sgn}[\delta_{xz,yz}(\bm{k})]}{\delta_{xy,yz}(\bm{k})} \right) \\
 0 
\end{pmatrix} , \label{eq:STO_g2_LS} \\
\bm{g}_{3}^{L}(\bm{k})&=  -2\lambda\gamma 
\begin{pmatrix}
 \sin k_y\left( \cfrac{1}{\delta_{xy,xz}(\bm{k})} \right) \\
 -\sin k_x \left( \cfrac{1}{\delta_{xy,yz}(\bm{k})}\right) \\ 
 0 
\end{pmatrix} . \label{eq:STO_g3_LS}
\end{align}
Figure \ref{fig:ASOC_L} shows the $\bm{k}$-dependence of the effective $g$-vector $\bm{g}_{\alpha}^{L}(\bm{k})$ for each Rashba split bands at $k_z=0$. 
We see that the unconventional Rashba spin-orbit splitting in the bulk STO [Fig. \ref{fig:STO_Rashba}] is well reproduced by the above perturbation analysis. 

According to Eqs. (\ref{eq:STO_g1_LS}), (\ref{eq:STO_g2_LS}), and (\ref{eq:STO_g3_LS}), the multiorbital effect is reflected in the ASOC through the energy difference $\delta_{ll'}(\bm{k})$ in two ways. 
One is the denominators, i.e., $\delta_{xy,yz}(\bm{k})$ and $\delta_{xy,xz}(\bm{k})$, and the other is the numerator, i.e., $1\pm\mathrm{sgn}[\delta_{xz,yz}(\bm{k})]$. 
The origin of the unconventional Rashba splitting in the upper band $\delta\mathcal{E}_{3}(\bm{k})$ is explained by the former multiorbital effect in the denominator of Eq. (\ref{eq:STO_g3_LS}) \cite{yanase2013electronic}. 
The magnitude of $\delta\mathcal{E}_{3}(\bm{k})$ is small on the line $\bm{k}\parallel[100]$ [Fig. \ref{fig:ASOC_L}(c)] since $\delta_{xy,yz}(\bm{k})$ is large and the $y$ component of the $g$-vector $\bm{g}_{3}^{L}(\bm{k})$ is small. 
On the other hand, a large $x$ component of $\bm{g}_{3}^{L}(\bm{k})$ appears upon moving slightly away from the $\Gamma$-X line because of the small value of its denominator $\delta_{xy,xz}(\bm{k})$. 
Thus, the spin-orbit splitting in the upper band is large at $\bm{k}$ slightly away from the $\Gamma$-X line as shown in Figs. \ref{fig:STO_Rashba}(a) and \ref{fig:ASOC_L}(c). 

The unconventional Rashba splitting in the lower and middle bands are explained by the combination of two multiorbital effects represented by the denominator and numerator of Eqs. (\ref{eq:STO_g1_LS}) and (\ref{eq:STO_g2_LS}). 
Since the denominators of Eqs. (\ref{eq:STO_g1_LS}) and (\ref{eq:STO_g2_LS}) are the same as those of Eq. (\ref{eq:STO_g3_LS}), the Rashba splitting different from that of the upper band originates from the numerator $1\pm\mathrm{sgn}[\delta_{xz,yz}(\bm{k})]$. 
The $\bm{k}$-dependence of $1\pm\mathrm{sgn}[\delta_{xz,yz}(\bm{k})]$ at $k_z=0$ is described as follows:  
\begin{align}
\eta_{\pm}(k_x,k_y) &\equiv 1\pm\mathrm{sgn}[\delta_{xz,yz}(k_x,k_y,k_z=0)] \nonumber\\
&=
\begin{cases}
    0 & (|k_x|\lessgtr|k_y|) \\
    1 & (|k_x|=|k_y|) \\
    2 & (|k_x|\gtrless|k_y|)
  \end{cases} . \label{eq:delta_eps}
\end{align}
Since $\eta_{-}(k_x,k_y)=0$ ($\eta_{+}(k_x,k_y)=0$) in $|k_x|>|k_y|$ ($|k_x|<|k_y|$), the $x$ ($y$) component of $\bm{g}_{1}^{L}(\bm{k})$ becomes zero. 
Thus, $\bm{g}_{1}^{L}(\bm{k})$ is parallel to the [100] or [010] axis in the most region of Brillouin zone, and rapidly rotates by $\pi/2$ when going across the line $|k_x|=|k_y|$ as shown in Figs. \ref{fig:STO_Rashba}(c) and \ref{fig:ASOC_L}(a). 
Figure \ref{fig:ASOC_L}(a) also shows that the spin-orbit splitting is maximized near the $\Gamma$-point, and rapidly decreases by increasing the distance from the $\Gamma$-point. 
From similar discussions we understand that $\bm{g}_2^L(\bm{k})$ is perpendicular to $\bm{g}_1^L(\bm{k})$ except for the line $|k_x|=|k_y|$.
Consequently, the Rashba splitting in the middle band becomes similar to that of the upper band, except the rapid $\pi/2$-rotation of the $g$-vector at $|k_x|=|k_y|$ [Figs. \ref{fig:STO_Rashba}(b) and \ref{fig:ASOC_L}(b)].

\subsection{\label{sec:J}Total angular momentum description}
Although the above perturbation analysis for the LS coupling reproduces many features of unconventional Rashba spin-orbit splitting in STO, it is not valid in the vicinity of the $\Gamma$-point. 
In particular, the disappearance of the spin-orbit splitting in the lower band near the $\Gamma$-point [Fig. \ref{fig:STO_Rashba}(c)] is not reproduced by the perturbation analysis [Fig. \ref{fig:ASOC_L}(a)]. 
This is because the wave function is appropriately labeled by the total angular momentum $J=L+S$ and the perturbation analysis breaks down. 
Therefore, it is desirable to derive the effective ASOC in the total angular momentum description. 
Generally speaking, the following analysis is valid for the strong spin-orbit coupling compared to other energy scales such as the band-width. 
Actually, we will see that it is valid only near the $\Gamma$-point. 
\begin{figure*}[tbp]
   \centering
   \includegraphics[width=180mm,clip]{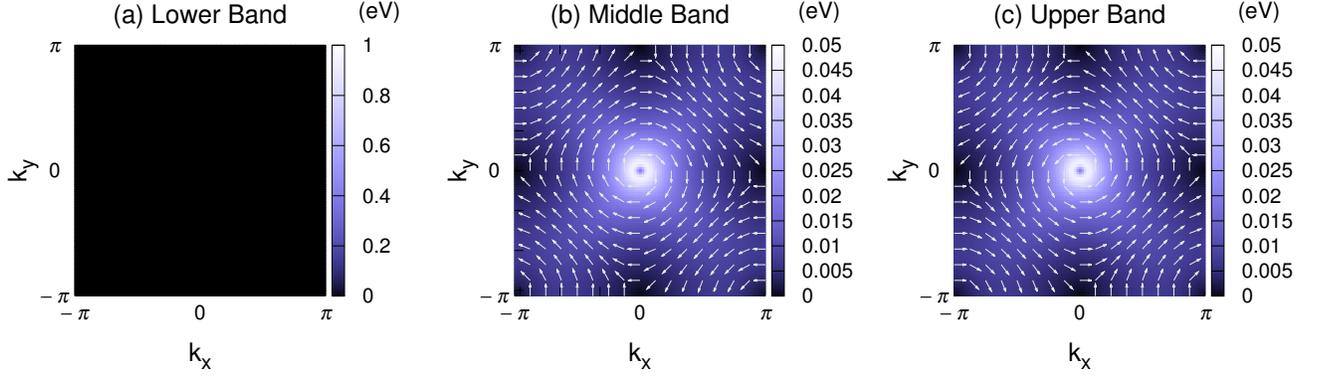}
\caption{The magnitude of the spin-orbit splitting, which is derived by the basis transformation to the total angular momentum space, in the (a) lower band, (b) middle band, and (c) upper band at $k_z=0$. The odd-parity hopping integral is set to $\gamma/t_1=0.105$. The arrows show the direction of the effective $g$-vector $\bm{g}_{\alpha}^{J}(\bm{k})$. 
\label{fig:ASOC_J} }
\end{figure*}

In the $t_{2g}$ orbital subspace, the orbital angular momentum can be treated as $L=1$. 
Thus, the total angular momentum $J=3/2$ or $J=1/2$ are obtained as a composition of the angular momentum $L=1$ and $S=1/2$. 
The basis wave functions in the total angular momentum space $\ket{J,M}$ are obtained as follows: 
\begin{align}
\ket{\frac{3}{2},\frac{3}{2}} 
&=-\frac{1}{\sqrt{2}}\Bigl(\ket{d_{yz},\uparrow}+i\ket{d_{xz},\uparrow}\Bigr), \label{eq:Q1} \\
\ket{\frac{3}{2},\frac{1}{2}} 
&= \frac{1}{\sqrt{6}}\Bigl(2\ket{d_{xy},\uparrow}-\ket{d_{yz},\downarrow}-i\ket{d_{xz},\downarrow} \Bigr) ,\\
\ket{\frac{3}{2},-\frac{1}{2}} 
&= \frac{1}{\sqrt{6}}\Bigl(\ket{d_{yz},\uparrow}-i\ket{d_{xz},\uparrow}+2\ket{d_{xy},\downarrow} \Bigr) , \\
\ket{\frac{3}{2},-\frac{3}{2}} 
&=\frac{1}{\sqrt{2}} \Bigl( \ket{d_{yz},\downarrow} - i \ket{d_{xz},\downarrow} \Bigr) , \label{eq:Q2} \\
\ket{\frac{1}{2},\frac{1}{2}} 
&= \frac{1}{\sqrt{3}}\Bigl(\ket{d_{xy},\uparrow}+\ket{d_{yz},\downarrow}+i\ket{d_{xz},\downarrow} \Bigr) , \\
\ket{\frac{1}{2},-\frac{1}{2}} 
&= \frac{1}{\sqrt{3}}\Bigl(\ket{d_{yz},\uparrow}-i\ket{d_{xz},\uparrow}-\ket{d_{xy},\downarrow} \Bigr), 
\end{align}
where $M=\pm3/2, \pm1/2$ is the total magnetic quantum number. 
Then, we carry out the $\bm{k}$-independent basis transformation for $\mathcal{H}_{0}+\mathcal{H}_{\rm pol}$ from the local $t_{2g}$ orbital space $\ket{t_{2g},\sigma}$ to the total angular momentum space $\ket{J,M}$. 
From the procedure similar to the previous subsection, the effective Hamiltonian is derived as 
\begin{align}
\tilde{\mathcal{H}}_{0}^{J} =& 
\sum_{\bm{k},\alpha,\sigma} 
\left(\varepsilon_{\alpha}^{J}({\bm k})-\mu\right) c_{{\bm k},\alpha \sigma}^{\dag} c_{{\bm k},\alpha \sigma} \nonumber\\
&+ \sum_{\bm{k},\alpha,\sigma,\sigma'} 
\bm{g}_{\alpha}^{J}(\bm{k}) \cdot \bm{\sigma}_{\sigma\sigma'} c_{{\bm k},\alpha \sigma}^{\dag} c_{{\bm k},\alpha \sigma'} .
\label{eq:H_eff_J}
\end{align}
Here, the renormalized energy dispersion $\varepsilon_{\alpha}^{J}({\bm k})$  are obtained as follows:
\begin{align}
 \varepsilon_{1}^J(\bm{k}) =& \frac{1}{2} \Bigl( \varepsilon_{yz}(\bm{k})+\varepsilon_{xz}(\bm{k})-\lambda \Bigr) , \\
\varepsilon_{2}^J(\bm{k}) =& \frac{1}{4} \left( \sum_{l} \varepsilon_{l}(\bm{k})+\lambda \right) - \sqrt{f^{J}(\bm{k})} , \\
\varepsilon_{3}^J(\bm{k}) =& \frac{1}{4} \left( \sum_{l} \varepsilon_{l}(\bm{k})+\lambda \right) + \sqrt{f^{J}(\bm{k})}, 
\end{align}
where  
\begin{align}
f^{J}(\bm{k}) =& 2\gamma^2\left(\sin^2 k_y + \sin^2 k_x \right) \nonumber\\ 
&+\left(\frac{\delta_{xy,yz}(\bm{k})+\delta_{xy,xz}(\bm{k})-\lambda}{4} \right)^2 .
\end{align}
The effective $g$-vector $\bm{g}_{\alpha}^{J}(\bm{k})$ are described as  
\begin{align}
\bm{g}_{1}^{J}(\bm{k}) &= \bm{0} ,\\
\bm{g}_{2}^{J}(\bm{k}) &= - \bm{g}_{3}^{J}(\bm{k}) = \frac{\lambda\gamma}{\sqrt{f^{J}(\bm{k})}} 
\begin{pmatrix}
 \sin k_y \\ -\sin k_x \\ 0
\end{pmatrix}. \label{eq:gvec_J} 
\end{align}
Interestingly, the spin-orbit spitting vanishes in the lower band, i.e., $\bm{g}_{1}^{J}(\bm{k})=\bm{0}$. 
The lower band is labeled by $J=3/2$ and $M=\pm3/2$ in the total angular momentum picture, and $\ket{3/2,\pm3/2}$ do not contain the $d_{xy}$ orbital [see Eqs. (\ref{eq:Q1}) and (\ref{eq:Q2})]. 
This means that the orbital parity mixing effect \cite{yanase2013electronic}, which is a necessary condition for the Rashba spin-orbit splitting, does not occur in the lower band. 
Therefore, Rashba splitting does not occur in the lower band, and indeed, the Rashba splitting in the full Hamiltonian disappears near the $\Gamma$-point [Fig. \ref{fig:STO_Rashba}(c)] where the total angular momentum description is valid. 
On the other hand, Eq. (\ref{eq:gvec_J}) shows that the magnitude of Rashba splitting in the middle band is finite and same as that in the upper band, although the sign of the $g$-vector is opposite. 
The $\bm{k}$-dependence of $\bm{g}_{2}^{J}(\bm{k})=-\bm{g}_{3}^{J}(\bm{k})$ is similar to that of the conventional Rashba ASOC with $\bm{g}(\bm{k})=\left(\sin k_y, -\sin k_x, 0\right)$, except that the magnitude of the spin-orbit splitting is maximized around the $\Gamma$-point [Fig. \ref{fig:ASOC_J}]. 
These momentum dependences are different from the results of numerical diagonalization [Fig. \ref{fig:STO_Rashba}]. 
This means that the perturbation analysis for the LS coupling is better at most $\bm{k}$-points in the Brillouin zone.

\nocite{*}

\providecommand{\noopsort}[1]{}\providecommand{\singleletter}[1]{#1}%

\end{document}